\documentclass[11pt]{article}
\usepackage{amsmath}
\usepackage{graphicx}

\parskip=\baselineskip
\renewcommand{\P}{{\cal P}}

\newcounter{head}
\def\thehead{\arabic{head}}
\def\head#1{\vskip\baselineskip\noindent\refstepcounter{head}
\textbf{\large\thehead. #1:~}}

\makeatletter
\def\subalign{\@ifnextchar[\subalign@{\subalign@[]}}
\def\subalign@[#1]{\subequations\ifx#1\void\else\eqlabel{#1}\fi\align}

\makeatother

\def\braket#1{\left\langle#1\right\rangle}

\def\P{{\cal P}}
\def\H{{\cal H}}

\def\det#1{\left\|#1\right\|}
\def\det#1{{{\rm det}\left[#1\right]}}

\def\deriv#1#2{\frac{\partial#1}{\partial#2}}
\def\xderiv#1#2{\frac{\partial^2#1}{\partial#2^2}}
\def\dderiv#1#2#3{\frac{\partial^2#1}{\partial#2\partial#3}}

\def\Order#1{{\cal O}\left(#1\right)}

\def\ie{\textit{i.e.}}

\def\r{{(r)}}
\def\e{{\rm e}}

\def\disc{{\rm disc}}
\def\cont{{\rm cont}}

\def\abs#1{\left|#1\right|}

\begin{document}

\begin{center}
\Large {\bf Products of Random Matrices}\\[3mm]
\normalsize{\bf A. D. Jackson and B. Lautrup}\\
The Niels Bohr Institute\\
Copenhagen, Denmark\\
{\bf P. Johansen}\\
The Institute of Computer Science\\
University of Copenhagen, Denmark\\
and\\
{\bf M. Nielsen}\\
The IT-university of Copenhagen, Denmark\\[3mm]
 \rm\today
\end{center}

\abstract{\it We derive analytic expressions for infinite products
of random $2\times2$ matrices. The determinant of the target
matrix is log-normally distributed, whereas the remainder  is a
surprisingly complicated function of a parameter characterizing
the norm of the matrix and a parameter characterizing its
skewness. The distribution may have importance as an uncommitted
prior in statistical image analysis.}

\head{Introduction}\label{sIntroduction}
Considerable  effort has been invested over the last half century
in determining the spectral properties of ensembles of matrices
with randomly chosen elements and in discovering the remarkably
broad applicability of these results to systems of physical
interest. In spite of a similarly rich set of potential
applications (e.g.\ in the statistical theory of Markov processes
and in various chaotic dynamical systems in classical physics),
the properties of products of random matrices have received
considerably less attention.  See ref.\,\cite{1} for a survey of
products of random matrices in statistics and ref.\,\cite{2} for a
review of physics applications.

The purpose of the present manuscript is to consider in some
detail the limit for $N\to\infty$  of the ensemble of matrices
\begin{align}\label{eProduct}
Y = \left( 1 + \sqrt{\frac\tau N}X_1 \right)\left( 1 +
\sqrt{\frac\tau N}X_2 \right)\cdots\left( 1 + \sqrt{\frac\tau
N}X_N \right)
\end{align}
where $\tau>0$ is a real parameter and the $X_n$ are real $d\times
d$ matrices with all elements drawn at random on a distribution of
zero mean and unit variance. If this distribution has compact
support, the probability that the matrix $Y$ should become
non-positive definite vanishes for $N\to\infty$. In one dimension,
$d=1$, it is well-known from the law of large numbers that $\log
Y$ has a Gaussian distribution, but because of the
non-commutativity of matrix products, the distribution is much
more complicated for $d\ge2$.

In this paper we derive some general properties  for the limiting
distribution $\P(Y)$ and determine it explicitly for $d=2$. In
section \ref{sDiffusion}  we establish a compact diffusion
equation for the distribution valid for any $d$. In section
\ref{sAverages} we derive a simple expression for any average over
the distribution, and we  show that the determinant $\det Y$ has a
log-normal distribution.  Sections \ref{sTwoDim} and
\ref{sSolution} will be devoted to the determination of the
explicit form of $\P$ for $d=2$.  We shall first write the
diffusion equation using an appropriate parameterization of $Y$.
The resulting partial differential equation will then be solved
subject to the boundary condition that $\P(Y)$ supports only the
identity matrix in the limit of $\tau\to0$. This explicit solution
will require new integrals involving Jacobi functions. The
derivation of these integrals will be given in the Appendix.

\head{The diffusion equation}\label{sDiffusion}
The normalized probability distribution  is (for given $N$ and
variable $\tau$)
\begin{align}
\P_N(Y,\tau)=\braket{\delta\left[Y-\prod_{n=1}^N \, \left( 1 +
\sqrt{\frac\tau N}X_n \right)\right]}_{X_1,\ldots,X_N}
\end{align}
where the integrand is a product of $\delta$-functions for each
matrix element of $Y$ and the average runs over all the random
matrices. Pealing off the $N$th factor in the product and using
only that the $X_n$ are statistically independent, we derive the
following exact recursion relation
\begin{align}\label{eRecursion}
\P_N(Y,\tau)=\braket{\det{1 + \sqrt{\frac\tau
N}X}^{-´d}\P_{N-1}\left[Y\left(1 + \sqrt{\frac\tau
N}X\right)^{-1},\tau\frac{N-1}N\right]}_X
\end{align}
where the average is over the $N$th random matrix, here renamed
$X$. The determinantal prefactor  of $\P_{N-1}$ is the Jacobi
determinant arising from the general matrix rule
\begin{align}
\delta[Y-ZM]=\frac{\delta[YM^{-1}-Z]}{\det{\deriv{(ZM)}Z}}
\end{align}
with $M=1+\sqrt{\tau/N}~X$. Since
\begin{align}
\deriv{(ZM)_{ij}}{Z_{k\ell}}=\delta_{ik}M_{\ell j}~,
\end{align}
the Jacobian is block diagonal with $d$ identical blocks, and the
prefactor follows.

The recursion relation \eqref{eRecursion} is of the Markovian type
with the initial distribution $\P_0(Y,\tau)=\delta[Y-1]$. It
converges for $N\to\infty$ under very general conditions (which we
shall not discuss here) towards a limiting distribution
$\P(y,\tau) =\lim_{N\to\infty}\P_N(y,\tau)$.  Expanding  the
recursion relation  to $\Order{1/N}$ and using the fact that all
the matrix elements of $X$ are statistically independent with zero
mean and unit variance,
\begin{align}
\braket{X_{ij}}_X=0
&&\braket{X_{ij}X_{kl}}_X=\delta_{ik}\delta_{jl}~,
\end{align}
we obtain to leading order
\begin{align*}
\P_N&=\P_{N-1}+\frac\tau N\left(-\deriv{\P_{N-1}}\tau+ \frac12 d^2
(d+1){\P_{N-1}}\right.\\&\quad\left. + (d+1) Y_{ij}
\deriv{\P_{N-1}}{Y_{ij}} + \frac12 Y_{ik} Y_{jk}
\dderiv{\P_{N-1}}{Y_{i\ell}}{Y_{j\ell}}\right)
\end{align*}
with implicit summation over all repeated indices. The assumed
convergence towards a limiting distribution requires the
expression in the parenthesis to vanish in the limit, so that
\begin{align}\label{eDiffusion}
\deriv\P\tau = \frac12 d^2 (d+1) \P + (d+1) Y_{ij}
\deriv\P{Y_{ij}} + \frac12 Y_{ik} Y_{jk}
\dderiv\P{Y_{i\ell}}{Y_{j\ell}}~.
\end{align}
This is  a diffusion equation of the Fokker-Planck type with
$\tau$ playing the role of time. It must be solved subject to the
initial condition that $\P(y,0)=\delta[Y-1]$.

Both the diffusion equation and the initial condition are
invariant with respect to an orthogonal transformation $Y\to
M^\top Y M$, where $M$ is an orthogonal matrix satisfying $M^\top
M=1$. Since the number of free parameters in an orthogonal
transformation is $\frac12 d(d-1)$, the number of ``dynamic''
variables in the distribution is $d^2-\frac12
d(d-1)=\frac12d(d+1)$. Since the distribution only has support for
$\det Y>0$, this number consists of $d$ independent eigenvalues
and $\frac12d(d-1)$ rotation angles in a singular value
decomposition.

For $d=1$ the solution to \eqref{eDiffusion} which approaches
$\delta[Y-1]$ for $\tau\to0$ is
\begin{align}\label{e1D}
\P_{d=1}(Y)=\frac1{Y\sqrt{2\pi\tau}}\exp\left[-\frac{(\log
Y+\tau/2)^2}{2\tau}\right]~.
\end{align}
As expected, it is a log-normal distribution.

\head{Averages}\label{sAverages}
Remarkably, equation \eqref{eDiffusion} may be written in the much
simper form
\begin{align}\label{eDiffusion1}
\deriv\P\tau =
\frac12\dderiv{(Y_{ik}Y_{jk}\P)}{Y_{i\ell}}{Y_{j\ell}}
\end{align}
without any explicit reference to $d$. Defining the average of a
function $f(Y)$ by
\begin{align}
\braket{f}=\int f(Y)\P(Y)\,dY
\end{align}
with $dY=\prod_{ij}dY_{ij}$, we obtain from \eqref{eDiffusion1}
\begin{align}\label{eAverage}
\deriv{\braket f}\tau=\frac12\braket{Y_{ik}Y_{jk}\dderiv
f{Y_{i\ell}}{Y_{j\ell}}}~.
\end{align}
This equation permits in principle the determination of the moment
of any product of matrix elements. The first two are found to be
\begin{align}
\braket{Y_{ij}}&=\delta_{ij}\\
\braket{Y_{ij}Y_{kl}}&=e^{\tau d}\delta_{ik}\delta_{jl}
\end{align}
The exponential growth of the averages with ``time'' $\tau$ is a
consequence of the multiplicative nature of the problem.

The determinant $D=\det Y$ is, according to the definition of the
product \eqref{eProduct}, an infinite product of random real
numbers that converge towards unity, and $\log D$ must  have a
Gaussian distribution according to the law of large numbers. Its
mean and variance are, however, different from those of the
one-dimensional distribution \eqref{e1D}. The distribution of the
determinant is also an average
\begin{align}
F(D)=\braket{\delta\bigl(D-\det Y\bigr)}~.
\end{align}
Using the fact that
\begin{align}
\deriv{\det Y}{Y_{ij}}=\det Y\,Y_{ji}^{-1}~,
\end{align}
we obtain  the following equation for $F$
\begin{align}\label{eDeterminantDiffusion}
\deriv F\tau=\frac 12d\xderiv{(D^2F)}D=d\left( F+2D\deriv
FD+\frac12D^2\xderiv F D\right)~.
\end{align}
Apart from the factor $d$ in front, this is identical to the
diffusion equation \eqref{eDiffusion1} in one dimension.
Consequently the determinant has a log-normal distribution
\begin{align}\label{eDeterminant}
F(D)=\frac1{D\sqrt{2\pi \tau d}}\exp\left[-\frac{(\log D+\tau
d/2)^2}{2\tau d}\right]~,
\end{align}
which is obtained from \eqref{e1D} by replacing $\tau$ by $\tau
d$. The distribution has support only for positive values of $D$.
It can be shown in general (and we shall demonstrate it explicitly
for $d=2$ below) that the distribution of the determinant
factorizes in $\P$.

\head{The case $d=2$}\label{sTwoDim}
The first non-trivial case is $d=2$ where  the general matrix is
first parameterized using a quaternion or 4-vector notation
\begin{align}
Y=\begin{pmatrix}Y_0+Y_3&Y_1-Y_2\\Y_1+Y_2&Y_0-Y_3\end{pmatrix}~.
\end{align}
In this representation the determinant becomes a metric with two
``space'' and two ``time'' dimensions
\begin{align}
D=Y_0^2-Y_1^2+Y_2^2-Y_3^2~.
\end{align}
The structure of this expression and the positivity of $D$ suggest
the following parameterization in terms of one imaginary and two
real angles
\begin{subalign}
Y_0&=\sqrt D\cosh\psi\cos\theta\\
Y_1&=\sqrt D\sinh\psi\cos\phi\\
Y_2&=\sqrt D\cosh\psi\sin\theta\\
Y_3&=\sqrt D\sinh\psi\sin\phi~.
\end{subalign}
The Jacobi determinant of the transformation from $\{ Y_0, Y_1,
Y_2, Y_3 \}$ to $\{D, \psi, \theta, \phi \}$ is simply
\begin{align}\label{eJacobi}
J \sim D \sinh{\psi} \cosh{\psi}~.
\end{align}
Orthogonal $2\times2$ matrices are generated by the matrix
$\left(\begin{smallmatrix}0&-1\\1&0\end{smallmatrix}\right)$,
which is associated with $Y_2$. Thus, an orthogonal transformation
rotates the angle $\phi$, and  $\P(Y,\tau)$  must be independent
of $\phi$ as indicated above.

In these variables the diffusion equation \eqref{eDiffusion}
simplifies to
\begin{align}
\deriv\P\tau&=6\P
+6D\deriv\P D+ D^2\xderiv\P D\nonumber\\
&+\frac14(1+\tanh^2\psi)\xderiv\P\theta
+\frac14(\tanh\psi+\coth\psi)\deriv\P\psi+\frac14\xderiv\P\psi~.
\end{align}
Taking into account the factor of $D$ in the Jacobi determinant,
we replace the original distribution $\P$ with the product of the
determinant distribution $F(D)$ given in \eqref{eDeterminant} and
an as yet unknown function of $\psi$ and $\theta$,
\begin{align}
\P=\frac1D F(D)G(\psi,\theta)~,
\end{align}
and find that $G$ satisfies the diffusion equation
\begin{align}\label{eDiffusion2}
\deriv G\tau=\frac14(1+\tanh^2\psi)\xderiv G\theta
+\frac14(\tanh\psi+\coth\psi)\deriv G\psi+\frac14\xderiv G\psi~.
\end{align}
The corresponding normalization integral is found from the Jacobi
determinant,
\begin{align}\label{eNorm}
\int_0^{2\pi}d\theta\int_0^\infty d\psi\, 2\sinh2\psi\,
G(\psi,\theta)=1~.
\end{align}
This normalization integrals \eqref{eNorm} suggests that it is
more convenient to employ still another variable
\begin{align}
z = \cosh{2
\psi}=\frac{Y_0^2+Y_1^2+Y_2^2+Y_3^2}{Y_0^2-Y_1^2+Y_2^2-Y_3^2}~.
\end{align}
With this variable the normalization integral takes the form
\begin{align}
\int_0^{2\pi}d\theta\int_1^\infty dz\, G(z,\theta)=1~,
\end{align}
and the diffusion equation \eqref{eDiffusion2} becomes
\begin{align}\label{eDiffusion3}
\deriv G\tau=\frac14\frac{2z}{z+1}\xderiv G\theta +2z\deriv
Gz+(z^2-1)\xderiv Gz~.
\end{align}
This equation must be solved with the boundary condition  that $\P
(Y)$ in the limit $\tau \to 0$ reduces to a product of delta
functions which select only the identity matrix. This evidently
requires $Y_0\to1$ and $Y_{1,2,3}\to0$ and, consequently, $D\to1$,
$z\to1$, and $\theta\to0$. Since $F(D)\to\delta(D-1)$, the initial
condition takes the form
\begin{align}\label{eInitial3}
G(z,\theta)\to \delta(z-1)\delta(\theta)\qquad(\tau\to0)~.
\end{align}
The limiting distribution should be approached from above (\ie\
from  $z>1$).

The form of the diffusion equation \eqref{eDiffusion3} reveals
that $G$ may naturally be expanded in  a Fourier series
\begin{align}
G(z,\theta)=\frac1{2\pi}\sum_{n=-\infty}^\infty G_n(z)
\e^{in\theta}
\end{align}
with coefficients that obey
\begin{align}\label{eDiffusion4}
\deriv{G_n}\tau=-\frac14n^2\frac{2z}{z+1}G_n +2z\deriv
{G_n}z+(z^2-1)\xderiv{G_n}z~.
\end{align}
For the special case $n=0$, we recognize Legendre's differential
operator on the right. The normalization condition only affects
$G_0$ and becomes
\begin{align}\label{eNorm1}
\int_0^\infty dz\, G_0(z)=1~.
\end{align}
The initial condition \eqref{eInitial3} implies that
\begin{align}\label{eInitial4}
G_n(z)\to\delta(z-1)\qquad(\tau\to0)
\end{align}
for all $n$.

\head{Explicit solution}\label{sSolution}
All that remains is to determine the angular functions $G_n(z)$.
One relatively simple way  is to use Sturm-Liouville theory, and
we  now outline the main steps in this procedure.

The differential operator (``Hamiltonian'') appearing on the right
hand side of eqn.\,\eqref{eDiffusion4} may be written
\begin{align}
\H=\deriv{}z(z^2-1)\deriv{}z-\frac{n^2}4\frac{2z}{z+1}~,
\end{align}
which shows that it is Hermitean.  Let the spectral  variable
(which denumerates the eigenvalues and may be both discrete and
continuous) be denoted $r$, and let $g_n^\r(z)$ be the
eigenfunction corresponding to the eigenvalue $\lambda_n^\r$,
\begin{align}
\H g_n^\r(z)=\lambda_n^\r g_n^\r(z)~.
\end{align}
The Hermiticity of $\H$ guarantees that the eigenvalues are real
and that the eigenfunctions are both orthogonal and complete on
the interval $1\le z<\infty$,
\begin{align}
\int_1^\infty dz\,
g_n^\r(z)g_n^{(r')}(z)&=\frac{\delta_{r,r'}}{\mu^\r_n}
\label{eNormalization}\\
\sum_r\mu^\r_n g_n^\r(z)g_n^\r(z') &=\delta(z-z')\label{eComplete}
\end{align}
with a suitable measure, $\mu_n^\r$.

The solution of the diffusion equation \eqref{eDiffusion4} with
initial condition \eqref{eInitial4} takes the form
\begin{align}
G_n(z,\tau)=\sum_r\mu^\r_n g_n^\r(1)
g_n^\r(z)\exp\left(\lambda_n^\r \tau\right)~.
\end{align}
In view of the completeness \eqref{eComplete}, these functions
indeed satisfy the initial conditions at $\tau=0$. The appearance
of $g_n^\r(1)$ in this expression requires the eigenfunctions to
be regular at $z=1$.

We now present the complete solution of the eigenvalue problem.
(Further details are given in the Appendix.) The eigenvalue
spectrum contains discrete values (for $n\ge2$) as well as a
continuum
\begin{align}
\lambda_n^{(r)}=
\begin{cases}
 -\frac12n^2-\frac14+\left(\frac{n+1}2-k\right)^2\quad
 &k=1,2,\ldots,\lfloor\frac n2\rfloor,\quad (n\ge2)\\
 -\frac12n^2-\frac14-t^2 &0\le t<\infty~.
\end{cases}
\end{align}
The properly normalized discrete eigenfunctions are  Jacobi
polynomials
\begin{align}
g_n^{(k)}=\sqrt{\frac{n+1}{2}-k} \, \left( \frac{1+z}{2}
\right)^{n/2} \, P_{-k}^{(0,n)} (z )~,
\end{align}
while the eigenfunctions in the continuum are   Jacobi functions
of complex index
\begin{align}
g_n^{(t)}=\left( \frac{1+z}{2}
\right)^{n/2}\,P_{-(n+1)/2+it}^{(0,n)} (z)
\end{align}
with the measure obtained from the integral \eqref{eNormalization}
as
\begin{align}\label{eNormalizationConstant}
\mu_n^{(t)}= \begin{cases}
t\tanh \pi t\quad&\text{$n$ even}\\
t\coth \pi t&\text{$n$ odd}~.
\end{cases}
\end{align}
The special case $n=0$ was stated without proof by Mehler in
1881~\cite{3}. The general case is proven in the Appendix.

\begin{figure}[t]
\includegraphics[width=0.9\hsize]{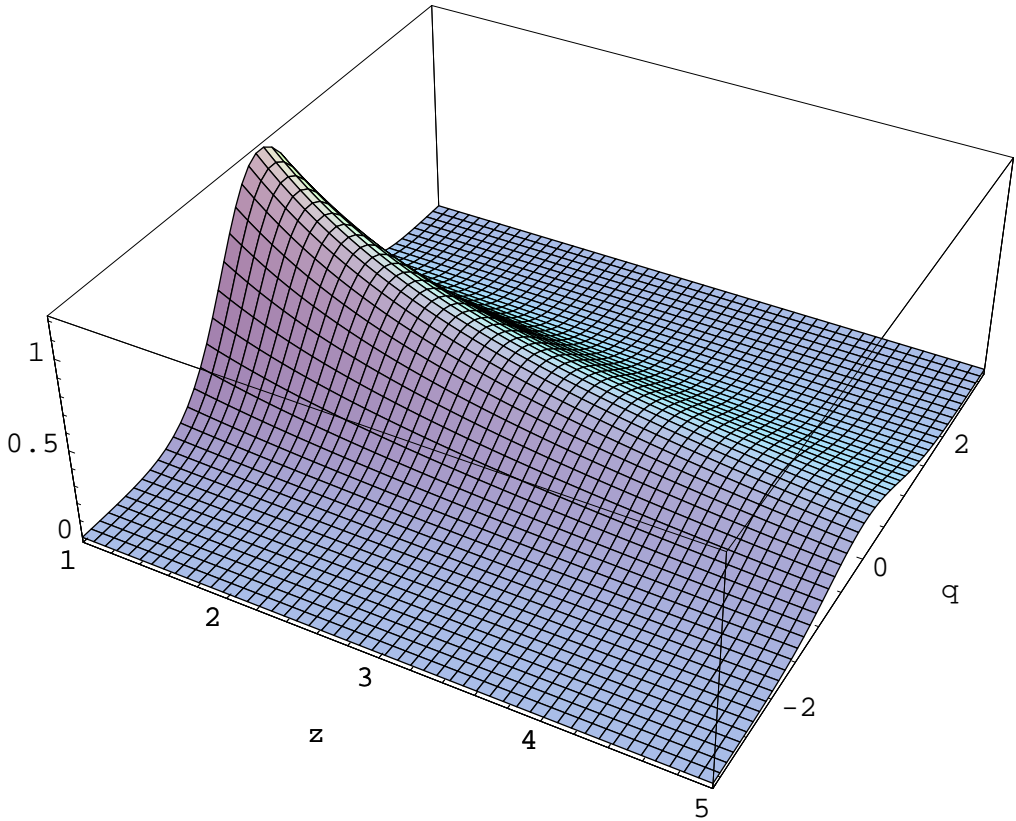}
\caption{\small\it Plot of $G(z,\theta)$ for $\tau=1$. Notice the
characteristic log-normal tapering of the ridge as a function of
$z$, and the nearly Gaussian distribution in $\theta$ around
$\theta=0$. }
 \label{fG}
\end{figure}

Since $g_n^{(t)}(1)=1$, the final solution becomes a simple
superposition of the discrete and continuous contributions
\begin{align}\label{eTotal}
G_n=G_n^\disc+G_n^\cont
\end{align}
where the discrete contribution (for $n\ge2$) is
\begin{align}\label{eDiscrete}
G_n^\disc(z,\tau)=\left( \frac{1+z}{2} \right)^{n/2}
\,\sum_{k=1}^{\lfloor n/2\rfloor} \, \left(\frac{n+1}{2}-k\right)
\, P_{-k}^{(0,n)} (z
)\e^{-\left(n^2/2+1/4-((n+1)/2-k)^2\right)\tau}~.
\end{align}
The continuous contribution is
\begin{align}\label{eContinuous}
G_n^\cont(z,\tau)=\left( \frac{1+z}{2} \right)^{n/2} \,
\int_0^{\infty} \, dt\,\mu_n(t)\, P_{-(n+1)/2+it}^{(0,n)} (z) \,
\e^{-(n^2/2+1/4+t^2)\tau}
\end{align}
with $\mu_n(t)$ given by \eqref{eNormalizationConstant}. Thus, we
arrive at the final result. The probability for drawing a given
$2\times2$ matrix $Y$ is
\begin{align}
\P(Y,\tau) = \frac{F(D)}{2\pi D} \, \,
\left(G_0(z,\tau)+2\sum_{n=1}^\infty G_n(z,\tau)\cos
n\theta\right)
\end{align}
with $F(D)$ given by eqn.\,\eqref{eDeterminant} and $G_n (z ,
\tau)$ given by eqns.\,(\ref{eTotal}--\ref{eContinuous}).  As
noted previously, the $G_n (z , \tau)$ are independent of the sign
of $n$ so that $\P$ is manifestly real. In fig. \ref{fG} the
function $G(z,\theta)$ (the expression in parenthesis) is plotted
for $\tau=1$.

\head{Conclusions}\label{sConclusions} We have analytically
derived the distribution of an infinite product of random
$2\times2$ matrices. In statistical image analysis, it may be used
as an uncommitted prior for morphing and warping \cite{4}, with
desirable properties not shared by the usual priors based on
elastic membranes. The distribution of such matrices   may be
evaluated numerically at a moderate cost in computer time and
converges reasonably fast because of the strong exponential
damping.

\newpage

\head{Appendix}
The Jacobi functions are related to the hypergeometric functions,
\begin{align} P_{-n/2-1/2+it}^{(0,n)} (z) =
{}_2F_1 \left(\frac{n+1}{2}  - it , \frac{n+1}{2} + it ; 1 ;
\frac{1-z}2 \right)
\end{align}
with $t$ real, and obey the orthogonality relation
\begin{align}\label{eNormalize1}
\int_1^{\infty} \, \left(\frac{1+z}2\right)^n \, dz \,
P_{-n/2-1/2+it}^{(0,n)} (z) \,
 P_{-n/2-1/2+it'}^{(0,n)} (z) =  \frac{\delta (t - t')}{\mu_n(t)}
\end{align}
In order to find $\mu_n(t)$ for arbitrary $n$, it is helpful to
consider the asymptotic form of these functions by using the
standard relation for hypergeometric functions
\begin{align}
F\left( a,b;c;z \right) & = & (1-z)^{-a} \,
\frac{\Gamma(c)\Gamma(b-a)}{\Gamma(b)\Gamma(c-a)} \,
F\left( a,c-b;a-b+1;\frac{1}{1-z} \right) + \nonumber \\
{} & {} & (1-z)^{-b} \,
\frac{\Gamma(c)\Gamma(a-b)}{\Gamma(a)\Gamma(c-b)} \, F\left(
b,c-a;b-a+1;\frac{1}{1-z} \right)~.
\end{align}
This form allows us to see that
\begin{align}
 P_{-n/2-1/2+it'}^{(0,n)} (z)  \to 2 \abs{A(t)} z^{-n/2-1/2}
\cos{(\phi_t + t \ln{z})}
\end{align}
as $z \to \infty$.  Here,
\begin{align}
A(t) = \frac{\Gamma(2it)}{\Gamma(n/2+1/2+it)\Gamma(-n/2+1/2+it)}\,
2^{n/2+1/2-it}~,
\end{align}
and $\phi_t$ is the phase of $A(t)$.  Using this asymptotic form,
we can perform the integral in eqn.\,\eqref{eNormalize1} by using
the variable $u = \log{z}$, adding a convergence factor of
$\exp{(-\mu u)}$, and finally taking the limit $\mu \to 0$.  The
result is simply
\begin{align}
\abs{A(t)}^2 \left[ \frac{2 \mu}{ \mu^2 + (t-t')^2}
\right]2^{-n}~.
\end{align}
The  factor in brackets is a familiar representation of $2 \pi
\delta (t - t')$ in the limit $\mu \to 0$. Standard relations for
the gamma function immediately yield eqn.\,\eqref{eNormalization}.
This confirms the results of Mehler~\cite{3} for the special case
$n=0$. The extension to $n > 0$ would appear to be new.


\begin{thebibliography}{99}

\bibitem{1}
Richard D. Gill and S{\o}ren Johansen, Ann.\,Statist.\ {\bf 18}
(1990) 1501.

\bibitem{2}
A. Crisanti, G. Paladin, and A. Vulpiani, {\em Products of Random
Matrices in Statistical Physics}, Springer-Verlag, Berlin, 1993.

\bibitem{3}
F. G. Mehler, Math.\,Ann.\ {\bf XVIII} (1881) 161.

\bibitem{4} Manuscript in preparation.

\end{thebibliography}
\end{document}